\documentstyle[11pt,epsfig]{article}
\title{\bf Compactification and signature transition
in Kaluza-Klein spinor cosmology}

\author{B. Vakili\thanks{email:
b-vakili@cc.sbu.ac.ir}, S. Jalalzadeh\thanks{email:
s-jalalzadeh@cc.sbu.ac.ir}
  and H. R. Sepangi\thanks{email:
hr-sepangi@cc.sbu.ac.ir}
\\ {\small Department of Physics, Shahid Beheshti University, Evin,
Tehran 19839, Iran}}
\begin{document}
\maketitle %\baselineskip 24pt

\begin{abstract}
We study the classical and quantum cosmology of a
$4+1$-dimensional space-time with a non-zero cosmological constant
coupled to a self interacting massive spinor field. We consider a
spatially flat Robertson-Walker universe with the usual scale
factor $R(t)$ and an internal scale factor $a(t)$ associated with
the extra dimension. For a free spinor field the resulting
equations admit exact solutions, whereas for a self interacting
spinor field one should resort to a numerical method for
exhibiting their  behavior. These solutions give rise to a
degenerate metric and exhibit signature transition from a
Euclidean to a Lorentzian domain. Such transitions suggest a
compactification mechanism for the internal and external scale
factors such that $a\sim R^{-1}$ in the Lorentzian region. The
corresponding quantum cosmology and the ensuing Wheeler-DeWitt
equation have exact solutions in the mini-superspace when the
spinor field is free, leading to wavepackets  undergoing signature
change. The question of stabilization of the extra dimension is
also discussed.
\vspace{5mm}\\
PACS umbers: 04.20.-q, 04.50.+h, 04.60.-m
\end{abstract}\pagebreak
%\tableofcontents

\section{Introdution}
Higher-dimensional cosmology resulting from the solutions of
higher-dimensional Einstein field equations begun to develop with
the works of Kaluza and Klein when they tried to geometrically
unify gravitational and electromagnetic interactions by
introducing an extra dimension in the space-time metric. Over a
long period of time this work has been the focus of an active area
of research \cite{1}. The extra dimension in Kaluza-Klein approach
has compact topology with compactification scale $l$ so that at
scales much larger than $l$ extra dimensions are not observable.
However, if one relaxes this condition by allowing the extra
dimensions to assume large sizes, interesting physical effects
appear. The birth of brane cosmology some few years ago has been
the focus of much attention in this regard  \cite{2,3}. The
assumption that our observable universe may be embedded in a
higher dimensional bulk has opened a new window in cosmology with
different approaches adopted in utilizing the size of extra
dimensions involved \cite{4}. A rather similar approach to models
with large extra dimensions comes from the space-time-matter
theory where our 4D world is embedded in a 5D bulk devoid of
matter \cite{5}. This theory was later shown to be basically
equivalent to the brane approach \cite{ponce}.

A question of interest in the classes of problems dealing with
higher dimensional cosmology is the mechanisms through which
compactification of extra dimensions can be achieved. One of the
common methods in this regard is to use mechanisms based on
signature transition in classical and quantum cosmology, first
addressed in the works of Hartle and Hawking \cite{6}. They argued
that quantum cosmology amplitudes should be expressed as the sum
of all compact Riemannian manifolds whose boundaries are located
at the signature changing hypersurface. This phenomenon has been
studied at the classical and quantum cosmology level by other
authors, see for example \cite{7}, \cite{D} and \cite{9}. There
are a number of works which support the idea that signature
transition would provide a mechanism for compactification in
higher dimensional cosmology \cite{10,11,12,13}. In \cite{10} a
higher dimensional FRW model with a positive cosmological constant
was considered with topology $S^3\times S^6$ as the spatial
section  and two scale factors $a_1$ and $a_2$. It was shown that
classical signature change induces compactification, that is,
drags the size of $S^6$ down and gives rise to a long-time
stability at an unobservable small scale. In \cite{11} the
compactification mechanism is studied for a $D+1$-dimensional
compact Kaluza-Klein cosmology with a negative cosmological
constant in which the matter field is either dust or coherent
excitations of a dilaton field. Another effort in this direction
can be found  in \cite{12} where the authors consider an empty
$4+1$-dimensional Kaluza-Klein cosmology with a negative
cosmological constant in which the external space is an open FRW
metric and the internal space contains a compact scale factor $a$.
It is then shown that signature transition induces
compactification on the scale factor $a$ in the Lorentzian domain,
dragging it to a small size of order $|\Lambda|^{1/2}$. These
efforts have mainly been made  on the assumption that either the
space is empty with a cosmological constant or filled with a
scalar field. The cases in which the matter source is a spinor
field have seldom been considered in the literature and it would
thus be of interest to employ such fields in this study.

In general, theories studying spinor fields coupled to gravity
result in Einstein-Dirac system which are somewhat difficult to
solve. The cosmological solutions of such systems have been
studied in a few cases by a number of authors \cite{Zh}. A general
discussion on the possibility that classical homogeneous spinor
fields can play the role of matter in cosmology can be found in
\cite{Ar}. The results of such a spinor cosmology  are notable
relative to the scalar field driven cosmology, particularly in
inflation scenarios \cite{Ob} and the prediction of cyclic
universes \cite{Ba}. In quantum cosmology, the Wheeler-DeWitt (WD)
equation for a universe filled with a scalar field is a
Schr\"{o}dinger like equation with a potential barrier \cite{Vi},
suggesting  a tunnelling mechanism for creation of the universe
from nothing. In contrast, in spinor quantum cosmology, the WD
equation becomes a Schr\"{o}dinger like equation with a potential
well and thus the creation of the universe can not be described by
such tunnelling procedures, for a discussion on this issue see
\cite{9}.

In this paper we consider a $4+1$-dimensional Kaluza-Klein
cosmology with a spatially flat Robertson-Walker metric, having
two dynamical variables, the usual external scale factor $R$ and
the internal scale factor $a$. The matter source in our model is a
Dirac spinor field which can be free or self-interacting. The
solutions of the resulting field equations show that they are
continuous functions passing smoothly from a Euclidean to a
Lorentzian domain through the signature changing hypersurface.
They also provide a mechanism for compactification of the scale
factors with different compactification scales, that is, of the
order $|\Lambda|^{-1/2}$ for $R$ and $|\Lambda|^{1/2}$ for $a$.
These results are in agreement with the results of \cite{12}. We
also touch upon the interesting question of stabilization of the
extra dimension in the case where the spinor field is self
interacting. Finally, the quantum cosmology of the model is
studied by presenting exact solutions to the WD equation in the
case of free spinor fields.
\section{The action}
We start by considering a cosmological model in which the
spacetime is assumed to be of spatially flat Robertson-Walker type
with one compact extra dimension as the internal space such that
the total bulk manifold has warped product topology
\begin{equation} \label{A}
M=M_{1+3}\times S^1.
\end{equation}
Adopting the chart $\{\beta,x,y,z,\rho\}$ where $\beta$ represents
the lapse function, $x$,$y$,$z$ denote the external space
coordinates and $\rho$ is the internal space coordinate, the
metric can be written as
\begin{equation} \label{B}
ds^2=-\beta d
\beta^2+R^{2}(\beta)\left(dx^2+dy^2+dz^2\right)+a^{2}(\beta)d\rho^2,
\end{equation}
where $R(\beta)$ and $a(\beta)$ are the scale factor of the
universe and the radius of the internal space respectively. From
(\ref{B}) it is clear that the sign of $\beta$ determines the
geometry, being Lorentzian if $\beta>0$ and Euclidean if
$\beta<0$. For $\beta>0$ the traditional cosmic time can be
recovered by the substitution $t=\frac{2}{3}\beta^{3/2}$ and
metric (\ref{B}) becomes
\begin{equation} \label{C}
ds^2=-dt^2+R^{2}(t)\left(dx^2+dy^2+dz^2\right)+a^{2}(t)d\rho^2,
\end{equation}
where $R(t)=R(\beta(t))$ and $a(t)=a(\beta(t))$ in the
$\{t,x,y,z,\rho\}$ chart. It should be noted that what we have
called the lapse function here is slightly different than the much
more popular ADM lapse function since this one enters the metric
as beta instead of $\beta^2$. As in \cite{7} we formulate our
equations in a region that does not include $\beta=0$ and seek
real solutions for $R$ and $a$ passing smoothly through the
$\beta=0$ hypersurface. The Ricci scalar corresponding to metric
(\ref{C}) is
\begin{equation} \label{D}
{\cal
R}=6\left[\frac{\ddot{R}}{R}+\frac{\dot{R}^2}{R^2}\right]+2\frac{\ddot{a}}{a}
+6\frac{\dot{R}}{R}\frac{\dot{a}}{a},
\end{equation}
where a dot represents differentiation with respect to $t$.

With this preliminary setup we write the action functional as
\begin{equation} \label{E}
{\cal S}=\int \sqrt{-g}L dtd^{3}x d\rho,
\end{equation}
where
\begin{equation} \label{F}
L=L_{grav}+L_{matt}.
\end{equation}
Here, $L_{grav}$ represents the Einstein-Hilbert Lagrangian for
the gravitational field with a cosmological constant $\Lambda$
\begin{equation} \label{G}
L_{grav}={\cal R}-\Lambda.
\end{equation}
The term $L_{matt}$ in (\ref{F}) denotes the Lagrangian for the
matter source, which we assume is a Dirac spinor field $\psi$. As
is well known the spinor field Lagrangian in curved spacetime is
given by
\begin{equation} \label{H}
L_{matt}=\frac{1}{2}\left[\bar{\psi}\gamma^{\mu}(\partial_{\mu}+
\Gamma_{\mu})\psi-\bar{\psi}(\overleftarrow{\partial_{\mu}}-
\Gamma_{\mu})\gamma^\mu\psi \right]-V(\bar{\psi},\psi),
\end{equation}
where $\gamma^{\mu}$ $(\mu=0,1,2,3,4)$ are the Dirac matrices
associated with the spacetime metric satisfying the five
dimensional Clifford algebra
$\{\gamma^{\mu},\gamma^{\nu}\}=2g^{\mu\nu}$, $\Gamma_{\mu}$ are
the spin connections and $V(\bar{\psi},\psi)$ is a potential
describing the interaction of the spinor field $\psi$ with itself.
The $\gamma^{\mu}$ matrices are related to the flat Dirac matrices
$\gamma^a$ through the vielbeins $e^a_\mu$ as follows
\begin{equation} \label{I}
\gamma^{\mu}=e^{\mu}_{a}\gamma^a,\hspace{.5cm}
\gamma_{\mu}=e^{a}_{\mu}\gamma_a.
\end{equation}
For metric (\ref{C}) the vielbeins can be easily obtained from
their definition, that is,
$g_{\mu\nu}=e^a_{\mu}e^b_{\nu}\eta_{ab}$ leading to
\begin{equation} \label{J}
e^a_{\mu}=\mbox{diag}(1,R,R,R,a),\hspace{.5cm}
e^{\mu}_a=\mbox{diag}(1,1/R,1/R,1/R,1/a).
\end{equation}
Also, the spin connections satisfy the relation
\begin{equation} \label{K}
\Gamma_{\mu}=\frac{1}{4}g_{\nu\lambda}(\partial_{\mu}e^{\lambda}_a+
\Gamma^{\lambda}_{\sigma\mu}e^{\sigma}_a)\gamma^{\nu}\gamma^a.
\end{equation}
Thus for line element (\ref{C}), use of (\ref{I}) and (\ref{J})
yields
\begin{equation} \label{L}
\Gamma_0=0,\hspace{.3cm}\Gamma_i=-\frac{\dot{R}}{2}\gamma^0\gamma^i,
\hspace{.3cm}i=1,2,3,\hspace{2mm}\mbox{and}\hspace{2mm}
\Gamma_4=-\frac{\dot{a}}{2}\gamma^0\gamma^4,
\end{equation}
where $\gamma^0$, $\gamma^i$ and $\gamma^4$ are the Dirac matrices
in the five dimensional Minkowski spacetime and we may always
adopt a suitable representation in which $\gamma^0$ is diagonal
with $(\gamma^0)^2=-1$. The final remark about the Lagrangian
(\ref{H}) is that for consistency of Einstein field equations with
a spinor field as the matter source in the background metric
(\ref{C}), the spinor field $\psi$ must depend on $t$ only, that
is $\psi=\psi(t)$ \cite{9}.

Now, substituting (\ref{G}) and (\ref{H}) into (\ref{E}) and
integrating over spatial dimensions results in an effective
Lagrangian ${\cal L}$ in the mini-superspace
$\{R,a,\psi,\bar{\psi}\}$, that is
\begin{equation} \label{M}
{\cal
L}=\frac{1}{2}Ra\dot{R}^2+\frac{1}{2}R^2\dot{R}\dot{a}+\frac{1}{6}\Lambda
R^3a+\frac{1}{12}R^3a\left[\bar{\psi}\gamma^{0}\dot{\psi}-\dot{\bar{\psi}}
\gamma^{0}\psi-2V(\bar{\psi},\psi)\right].
\end{equation}
\section{Field equations}
Variation of Lagrangian (\ref{M}) with respect to $\bar{\psi}$,
$\psi$, $R$ and $a$ yields the equation of  motion of the spinor
field and the Einstein field equations
\begin{equation} \label{N}
\dot{\psi}+\left(\frac{3}{2}\frac{\dot{R}}{R}+
\frac{1}{2}\frac{\dot{a}}{a}\right)\psi+\gamma^0\frac{\partial
V}{\partial\bar{\psi}}=0,
\end{equation}
\begin{equation} \label{O}
\dot{\bar{\psi}}+\left(\frac{3}{2}\frac{\dot{R}}{R}+
\frac{1}{2}\frac{\dot{a}}{a}\right)\bar{\psi}-\frac{\partial
V}{\partial \psi}\gamma^0=0,
\end{equation}
\begin{equation} \label{P}
2\frac{\ddot{R}}{R}+\frac{\dot{R}^2}{R^2}+
2\frac{\dot{R}}{R}\frac{\dot{a}}{a}-\Lambda=
\frac{1}{2}\left(\bar{\psi}\frac{\partial V}{\partial
\bar{\psi}}+\frac{\partial V}{\partial
\psi}\psi\right)-V(\bar{\psi},\psi),
\end{equation}
\begin{equation} \label{Q}
3\frac{\dot{R}^2}{R^2}+3\frac{\ddot{R}}{R}-\Lambda=
\frac{1}{2}\left(\bar{\psi}\frac{\partial V}{\partial
\bar{\psi}}+\frac{\partial V}{\partial
\psi}\psi\right)-V(\bar{\psi},\psi).
\end{equation}
The ``zero energy condition" also reads
\begin{equation} \label{R}
{\cal H}=\frac{\partial {\cal L}}{\partial
\dot{R}}\dot{R}+\frac{\partial {\cal L}}{\partial
\dot{a}}\dot{a}+\frac{\partial {\cal L}}{\partial
\dot{\psi}}\dot{\psi}+\dot{\bar{\psi}}\frac{\partial {\cal
L}}{\partial \dot{\bar{\psi}}}-{\cal L}=0,
\end{equation}
yielding the constraint equation as
\begin{equation} \label{S}
3\frac{\dot{R}^2}{R^2}+3\frac{\dot{R}}{R}\frac{\dot{a}}{a}-
\Lambda=-V(\bar{\psi},\psi).
\end{equation}
It is clear that the right hand sides of equations (\ref{P}),
(\ref{Q}) and (\ref{S}) represent the components of the
energy-momentum tensor where the energy density of the spinor
field is given by
\begin{equation} \label{T}
\rho=T_{00}=-V(\bar{\psi},\psi).
\end{equation}

To make Lagrangian (\ref{M}) manageable, consider the following
change of variables
\begin{equation} \label{U}
R=\left(u+v\right)^{1/2},\hspace{.5cm}
a=\left(u+v\right)^{-1/2}\left(u-v\right).
\end{equation}
In terms of the new variables $(u,v)$, Lagrangian (\ref{M}) takes
the form
\begin{equation} \label{V}
{\cal L}=\frac{1}{4}\left(\dot{u}^2-\dot{v}^2\right)+
\frac{1}{6}\Lambda\left(u^2-v^2\right)+\frac{1}{12}
\left(u^2-v^2\right)\left[\bar{\psi}\gamma^{0}\dot{\psi}-
\dot{\bar{\psi}}\gamma^{0}\psi-2V(\bar{\psi},\psi)\right],
\end{equation}
with the energy constraint (\ref{S}) becoming
\begin{equation} \label{W}
{\cal H}=\frac{1}{4}\left(\dot{u}^2-\dot{v}^2\right)-
\frac{1}{6}\Lambda\left(u^2-v^2\right)+
\frac{1}{6}\left(u^2-v^2\right)V(\bar{\psi},\psi)=0.
\end{equation}
In terms of the new variables, the field equations
(\ref{N})-(\ref{Q})  can be obtained by the variation of
Lagrangian (\ref{V}) with respect to its dynamical variables
$(\bar{\psi},\psi,u,v)$ with the result
\begin{equation} \label{X}
\dot{\psi}+\frac{u\dot{u}-v\dot{v}}{u^2-v^2}\psi+\gamma^0\frac{\partial
V}{\partial \bar{\psi}}=0,
\end{equation}
\begin{equation}\label{Y}
\dot{\bar{\psi}}+\frac{u\dot{u}-v\dot{v}}{u^2-v^2}\bar{\psi}-\frac{\partial
V}{\partial \psi}\gamma^0=0,
\end{equation}
\begin{equation}\label{Z}
\ddot{u}-\frac{2}{3}\Lambda
u-\frac{1}{3}u\left(\bar{\psi}\frac{\partial V}{\partial
\bar{\psi}}+\frac{\partial V}{\partial \psi}\psi-2V\right)=0,
\end{equation}
\begin{equation}\label{a}
\ddot{v}-\frac{2}{3}\Lambda
v-\frac{1}{3}v\left(\bar{\psi}\frac{\partial V}{\partial
\bar{\psi}}+\frac{\partial V}{\partial \psi}\psi-2V\right)=0.
\end{equation}
Our goal would now be to concentrate on the solutions of these
equations for various forms of the potential $V$.
\section{Solutions}
Up to this point the cosmological model has been rather general.
However, motivated by the desire to find suitable smooth functions
$R(t)$, $a(t)$, $\psi(t)$ and in particular to stabilize the
internal degree of freedom, a suitable potential
$V(\bar{\psi},\psi)$ should be chosen. First, we assume that this
potential has the property
\begin{equation}\label{b}
\bar{\psi}\gamma^0\frac{\partial V}{\partial
\bar{\psi}}-\frac{\partial V}{\partial \psi}\gamma^0 \psi=0.
\end{equation}
This is not a severe restriction on the form of the potentials as
most of the potentials in use would satisfy the above condition.
With this property in mind, we can integrate equations (\ref{X})
and (\ref{Y}) to obtain
\begin{equation}\label{c}
\bar{\psi}\psi=\frac{{\cal C}}{u^2-v^2}=\frac{{\cal C}}{R^3a},
\end{equation}
where ${\cal C}$ is an integrating constant and we have used
transformations (\ref{U}).

In general the potential $V(\bar{\psi},\psi)$ should describe a
physical self-interacting spinor field and is usually an invariant
function constructed from the spinor $\psi$ and its adjoint
$\bar{\psi}$. Some of the common forms for $V$ are:
$V(\bar{\psi},\psi)=m\bar{\psi}\psi$ representing a free spinor
field of mass $m$,
$V(\bar{\psi},\psi)=m\bar{\psi}\psi+J^{\mu}J_{\mu}$ where
$J^{\mu}=\bar{\psi}\gamma^{\mu}\psi$ and known as the Thirring
model, $V(\bar{\psi}\psi)=m\bar{\psi}\psi+\lambda
(\bar{\psi}\psi)^{n}$ and so on. In what follows we focus
attention on two cases, the free   and   self-interacting spinor
field potentials with the latter having a term of the form
$(\bar{\psi}\psi)^{n}$.
\subsection{Free spinor field}
This is the simplest case in which the potential has the form
$V(\bar{\psi},\psi)=m\bar{\psi}\psi$. Use of equations (\ref{T})
and (\ref{c}) results in obtaining the energy density of the
spinor field
\begin{equation}\label{d}
\rho=-\frac{m {\cal C}}{R^3a}.
\end{equation}
The condition $\rho>0$ demands that the integrating constant
${\cal C}$ be negative and as such, we choose it to be $-1$. For
the initial conditions we take $\dot{u}(0)=\dot{v}(0)=0$, whose
relevance  will be discussed in the next section. Integrating
equations (\ref{Z}) and (\ref{a}), one is led to
\begin{equation}\label{e}
u(t)={\cal
A}\cos\left(\sqrt{-\frac{2}{3}\Lambda}t\right),\hspace{.5cm}
v(t)={\cal B}\cos\left(\sqrt{-\frac{2}{3}\Lambda}t\right),
\end{equation}
where ${\cal A}$ and ${\cal B}$ are two integrating constants.
Now, these solutions must satisfy the ``zero energy condition,"
equation (\ref{W}). Thus, substitution of equations (\ref{e}) into
(\ref{W}) gives a relation between ${\cal A}$ and ${\cal B}$
\begin{equation}\label{f}
{\cal A}^2-{\cal B}^2=\frac{m}{-\Lambda}.
\end{equation}
If one chooses the constants ${\cal A}$ and ${\cal B}$ such that
${\cal A}-{\cal B}=1$, then  one finds the following solutions in
terms of $\beta$
\begin{equation}\label{g}
R(\beta)=\left(\frac{m}{-\Lambda}\right)^{1/2}\cos^{1/2}
\left(\frac{2}{3}\sqrt{-\frac{2}{3}\Lambda}\beta^{3/2}\right),
\end{equation}
\begin{equation}\label{h}
a(\beta)=\left(\frac{-\Lambda}{m}\right)^{1/2}\cos^{1/2}
\left(\frac{2}{3}\sqrt{-\frac{2}{3}\Lambda}\beta^{3/2}\right).
\end{equation}
Also, the energy density and Ricci scalar are obtained as
\begin{equation} \label{i}
\rho=\frac{-\Lambda}{\cos^{2}\left(\frac{2}{3}
\sqrt{-\frac{2}{3}\Lambda}\beta^{3/2}\right)},\hspace{.5cm} {\cal
R}=-\Lambda\left[\tan^{2}\left(\frac{2}{3}\sqrt{-\frac{2}{3}
\Lambda}\beta^{3/2}\right)-3\right].
\end{equation}
The classical solutions (\ref{g}) and (\ref{h}) describe a
Kaluza-Klein universe with a cosmological constant filled with a
free spinor field. When $\Lambda<0$, both scale factors $R$ and
$a$ are unbounded in the Euclidean region $\beta<0$, passing
continuously through $\beta=0$ and exhibiting bounded oscillations
in the Lorentzian region $\beta>0$. For $|\Lambda|\gg 0$, these
solutions  give rise to a small scale factor $R$ passing smoothly
from Euclidean to Lorentzian regions. The scale factor $a$ will be
very large in both regions compared to the scale factor $R$ and
passes through $\beta=0$ smoothly. The scale factor $R$ is thus
compactified to a small size of order $|\Lambda|^{-1/2}$ in the
Lorentzian domain. Taking $|\Lambda|\simeq 0$, the scale factor
$a(\beta)$ becomes very small in comparison to $R(\beta)$. As it
is clear, it passes through $\beta=0$ continuously with a very
small value $a(0)=\left(\frac{|\Lambda|}{m}\right)^{1/2}$ and
oscillates for $\beta>0$ with amplitude
$\left(\frac{|\Lambda|}{m}\right)^{1/2}$. One of the interesting
features in this case is that signature transition can induce
compactification on the scale factor $a(\beta)$ in the Lorentzian
region, dragging it to a small size of order $|\Lambda|^{1/2}$.
The first zero of the oscillatory functions appearing in (\ref{g})
and (\ref{h}) in the $\beta>0$ region occurs at
\begin{equation}\label{j}
\beta_{0}=\left(\frac{3}{4}\pi\sqrt{\frac{3}{-2\Lambda}}\right)^{2/3}.
\end{equation}
From (\ref{j}) it is seen that for a small value of the
cosmological constant (large $\beta_0$) we have an extended
Lorentzian region $0\leq \beta \leq \beta_0$ which would
correspond to a Kaluza-Klein  cosmology with a large scale factor
$R$ and a stable compactified internal scale factor $a$. Note that
for $|\Lambda|\sim 10^{-56}\mbox{cm}^{-2}$ we have $\beta_{0}\geq$
\textit{present age of the universe}. These results are completely
in agreement with \cite{12} in which an empty Kaluza-Klein
universe with negative curvature $(k=-1)$ was investigated.
However, it is worth noting that for $k=0$ this model becomes
singular due to the appearance of $k$ in the denominator of the
scale factor $R$. In the present discussion our model is free of
such singularities for $k=0$.

When $\Lambda>0$, both scale factors $R(\beta)$ and $a(\beta)$
have oscillatory behavior in the Euclidean region passing smoothly
through $\beta=0$ and are unbounded in the Lorentzian domain.
Since signature transition does not lead to compactification and
the energy density of the spinor field becomes negative, this case
is not of interest in our present investigation.
\subsection{Self-interacting spinor field}
In this section we consider a self interacting potential which we
have chosen to be of the following form
\begin{equation}\label{k}
V(\bar{\psi},\psi)=m\bar{\psi}\psi+\lambda(\bar{\psi}\psi)^n,
\end{equation}
where $\lambda$ is a coupling constant and $n>0$. Since this
potential has the property represented by equation (\ref{b}),
integrating equations (\ref{X}) and (\ref{Y}) again yields
\begin{equation}\label{l}
\bar{\psi}\psi=\frac{{\cal D}}{u^2-v^2},
\end{equation}
where ${\cal D}$ is a constant which is not necessarily equal to
${\cal C}$ in (\ref{d}). Using this result in equations (\ref{Z})
and (\ref{a}) one obtains the following dynamical equations for
$u$ and $v$ in terms of the evolution variable $\beta$
\begin{equation}\label{m}
u^{''}=\frac{1}{2}\frac{1}{\beta}u^{'}-\frac{3}{4}k\beta u
(u^2-v^2)^{-n}+\frac{2}{3}\Lambda\beta u,
\end{equation}
\begin{equation}\label{n}
v^{''}=\frac{1}{2}\frac{1}{\beta}v^{'}-\frac{3}{4}k\beta v
(u^2-v^2)^{-n}+\frac{2}{3}\Lambda\beta v,
\end{equation}
where $\frac{3}{4}k=\frac{2}{3}(1-n)\lambda {\cal D}^n$ and a
prime represents differentiation with respect to $\beta$. Note
that $k$ should not be confused with curvature here. Also the zero
energy constraint (\ref{W}) takes the form
\begin{equation}\label{o}
-4{\cal
H}=\frac{1}{\beta}(-u{'}^2+v{'}^2)-\frac{9}{4}k(u^2-v^2)^{1-n}
+\frac{2}{3}\Lambda (u^2-v^2)-\frac{2}{3}m{\cal D}=0.
\end{equation}
The coupled equations (\ref{m}) and (\ref{n}) should now be solved
and since their solutions must automatically satisfy the energy
constraint, equation (\ref{o}) becomes only a restriction on the
initial conditions. Note that equations (\ref{m}) and (\ref{n})
are a system of coupled non-linear differential equations with
{\it moving singularities} \cite{C} at the critical values
$\beta_c$ at which $u(\beta_c)=\pm v(\beta_c)$. In terms of the
original variables we have $a(\beta_c)=0$, $R(\beta_c)=0$ and
$a(\beta_c)=\infty$. These equations do not have closed form
solutions. However, such system of equations have been
investigated  in \cite{D} and \cite{C} where analytic solutions
near $\beta=0$ are obtained and a numerical method to find the
solutions for the full range of $\beta$ is presented.

\begin{figure}
\centerline{\begin{tabular}{ccc}
\epsfig{figure=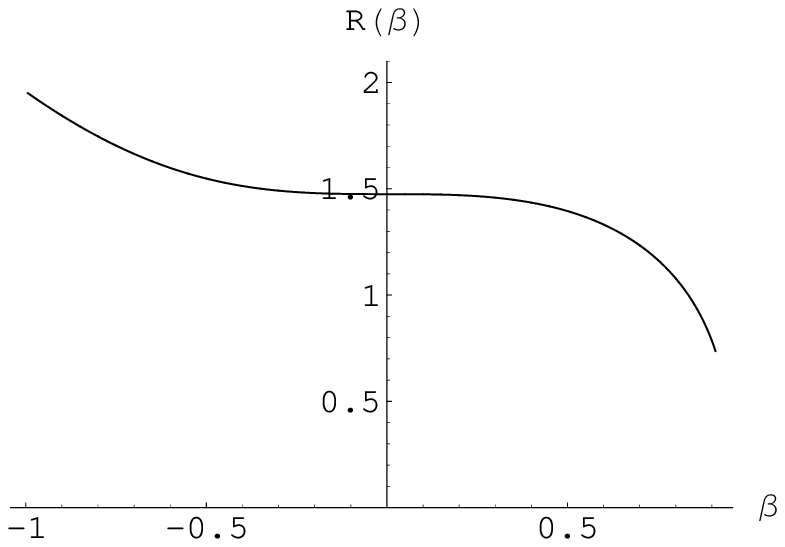,width=8cm}
 &\hspace{2.cm}&
\epsfig{figure=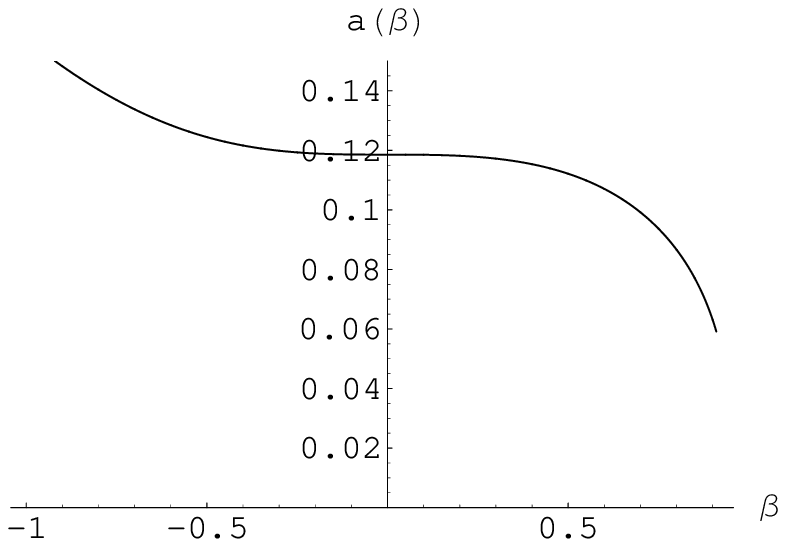,width=8cm}
\end{tabular}  }
\caption{\footnotesize Left, the scale factor $R$ for a self
interacting spinor field and right, the internal scale factor $a$
for the same field, both for $k=1$ and $n=\frac{4}{3}$ with
$\frac{2}{3}\Lambda=-1$ and $\frac{2}{3}m{\cal D}=1$.}
\label{fig1}
\end{figure}

For the problem at hand, in order to have well behaved solutions
closed to $\beta=0$, the first term in equation (\ref{o}) imposes
that $u'(\beta)\sim \beta^r$ and $v'(\beta)\sim \beta^p$, where
$r,p\geq \frac{1}{2}$ or $|u'(0)|=|v'(0)|$. On the other hand the
first terms on the right hand side of equations (\ref{m}) and
(\ref{n}) imply  that near $\beta=0$ these equations admit
solutions of the form $u'(\beta)\sim \beta^{1/2}$ and
$v'(\beta)\sim \beta^{1/2}$ which impose a more sever restriction
on the initial conditions. Also, they are not real or $C^2$ across
$\beta=0$. It is easy to show that regular solutions close to
$\beta=0$ are of the form \cite{D}
\begin{equation}\label{p}
u(\beta)=A\beta^3+u_0 \hspace{.3cm}\mbox{where} \hspace{.3cm}
A=-\frac{2}{9}\left[\frac{3}{4}\frac{ku_0}{(u_0^2-v_0^2)^n}-\frac{2}{3}\Lambda
u_0\right],
\end{equation}
\begin{equation}\label{q}
v(\beta)=B\beta^3+v_0 \hspace{.3cm}\mbox{where} \hspace{.3cm}
B=-\frac{2}{9}\left[\frac{3}{4}\frac{kv_0}{(u_0^2-v_0^2)^n}-\frac{2}{3}\Lambda
v_0\right],
\end{equation}
where $u_0=u(0)$ and $v_0=v(0)$. Thus the following initial
condition must be satisfied
\begin{equation}\label{r}
u'(0)=v'(0)=0 \hspace{.3cm} \mbox{and} \hspace{.3cm}
u''(0)=v''(0)=0.
\end{equation}
The above initial conditions are those we used for finding
solutions (\ref{e}). One should note that since we have a system
of coupled second order differential equations, the relations
$u''(0)=v''(0)=0$ are not initial conditions but rather
consistency checks. Therefore the initial values for functions $u$
and $v$ should now satisfy equation (\ref{o})
\begin{equation}\label{s}
\frac{9}{4}k(u_0^2-v_0^2)^{1-n}-\frac{2}{3}\Lambda
(u_0^2-v_0^2)+\frac{2}{3}m {\cal D}=0.
\end{equation}
Equation (\ref{s}) shows an implicit relation from which one can
find the possible initial values for $u$ and $v$. What remains is
the solution of the system of equations (\ref{m}) and (\ref{n})
with the initial conditions (\ref{s}). As mentioned before this
system cannot not be solved analytically and one may use the same
numerical method as described in \cite{D,C}.  The results for the
scale factors $R$ and $a$ are shown in figure \ref{fig1} for
typical values of the parameters. As it is clear from the figure,
both external scale factor $R$ and internal scale factor $a$ are
unbounded functions in the Euclidean region $\beta<0$ passing
smoothly through the signature changing hypersurface $\beta=0$ and
are bounded in the Lorentzian region $\beta>0$. Note that as in
the free spinor field case, signature transition induces
compactification on the scale factors but the compactification
scale for the internal space is much smaller than that for the
external space. Indeed equations (\ref{g}), (\ref{h}) and figure
\ref{fig1} show that in the presence of a free or self interacting
spinor field the corresponding Kaluza-Klein cosmology has a large
external scale factor in the Lorentzian domain with a size of
order $|\Lambda| ^{-1/2}$ together with a small compact internal
space having a radius of order $|\Lambda|^{1/2}$.
\section{Stabilization}
A problem of great importance in theories dealing with extra
dimensions is the question of their stabilization which we are now
going to address. In general, as it was shown in \cite{Ark},
stabilization of the extra dimension is related to the properties
of the potential appearing in the theory. From Lagrangian
(\ref{M}) we can write the potential as
\begin{equation} \label{1}
U=-\frac{1}{6}\Lambda R^3 a +\frac{1}{6}R^3 a V(\bar{\psi},\psi).
\end{equation}
Taking $V(\bar{\psi},\psi)=m\bar{\psi}\psi+\lambda
(\bar{\psi}\psi)^n$ and writing $\bar{\psi}\psi={\cal D}/R^3a$
from equation (\ref{l}), the potential takes the form
\begin{equation}\label{2}
U=-\frac{1}{6}\Lambda R^3 a +\frac{1}{6}\lambda {\cal D}^n
(R^3a)^{1-n}+\frac{1}{6}m {\cal D},
\end{equation}
whose  minimum occurs when $\frac{\partial U}{\partial R}|_{R_0}=0
$ and $\frac{\partial U}{\partial a}|_{a_0}=0$, yielding
\begin{equation} \label{3}
a_0\sim \left[\lambda {\cal
D}^{n}(n-1)|\Lambda|^{3n/2-1}\right]^{1/n},
\end{equation}
where we have used the approximation $R_0 \sim |\Lambda|^{-1/2}$
resulting from our classical solution (\ref{g}). Also, the second
derivative of $U$ with respect to $a$ becomes
\begin{equation}\label{4}
U''(a_0)=n(n-1)\lambda {\cal D}^n R_0^{3-3n}a_0^{-n-1},
\end{equation}
which for $n>1$ renders the potential minimum at $a_0$. Let us now
expand Lagrangian (\ref{M}) around this minimum as
$a(t)=a_0+\delta a$. Defining $\delta=\delta a/a_0$, it is easy to
show that $\delta=\delta_0 e^{i\omega t}$ where $\omega^2\propto
U''(a_0)>0$  \cite{Ark}. One should note that if the spinor field
is free, {\it  i.e.}  $\lambda=0$, then the potential does not
have a minimum and the internal  degree of freedom $a(t)$ is not
stable. This is also clear from  the classical solution (\ref{h})
which also shows that the internal scale factor oscillates between
zero and a finite size. In summery, having a stable internal space
is only possible in this model if the spinor field is  self
interacting with $n>1$ in equation (\ref{k}).

\section{Quantum cosmology}
We now focus attention on the  study of the quantum cosmology of
the model described above. We start by writing the Wheeler-DeWitt
equation from Hamiltonian (\ref{W}). As we saw earlier, a
self-interacting spinor field results in equations which are
complicated and difficult to solve. One would therefore expect the
corresponding quantum cosmology to become equally complicated.
However, this is not the case when we have a free spinor field.
This motivates us to concentrate on the quantum cosmology
corresponding to the classical solutions represented by equations
(\ref{g}) and (\ref{h}) which can be cast into an
oscillator-ghost-oscillator system whose solutions are well known.
The WD equation resulting from Hamiltonian (\ref{W}) with
$V(\bar{\psi},\psi)=m\bar{\psi}\psi$ and equation (\ref{c}) can be
written as
\begin{equation}\label{t}
{\cal H}\Psi(u,v)=\left[\frac{\partial^2}{\partial
u^2}-\frac{\partial^2}{\partial
v^2}-(u^2-v^2)\omega^2-\frac{1}{6}m\right]\Psi(u,v)=0,
\end{equation}
where $\omega^2=-\frac{1}{6}\Lambda$. This equation is separable
in terms of the mini-superspace variables and a solution can be
written as
\begin{equation}\label{u}
\Phi_{n_1,n_2}(u,v)=\alpha_{n_1}(u)\beta_{n_2}(v) \hspace{.5cm}
n_1,n_2=0,1,2,...
\end{equation}
where
\begin{equation}\label{v}
\alpha_n(u)=\left(\frac{\omega}{\pi}\right)^{1/4}\frac{1}{\sqrt{2^n
n!}}e^{-\omega u^2/2}H_n(\sqrt{\omega}u),
\end{equation}
\begin{equation}\label{w}
\beta_n(v)=\left(\frac{\omega}{\pi}\right)^{1/4}\frac{1}{\sqrt{2^n
n!}}e^{-\omega v^2/2}H_n(\sqrt{\omega}v),
\end{equation}
and $H_n(x)$ is a Hermite polynomial. The zero energy constraint
${\cal H}=0$ then yields
\begin{equation}\label{x}
n_2-n_1=\frac{1}{12}\frac{m}{\omega}.
\end{equation}
Note that since equation (\ref{d}) shows $m$ as the total energy
of the spinor field, this equation represents a quantization
condition of the spinor field energy. The set
$\{\Phi_{n_1,n_2}(u,v)\}$ is a base for the Hilbert space of the
measurable square-integrable functions on $R^2$ with the usual
inner product
\begin{equation}\label{y}
\int \Phi_{n_1,n_2}(u,v)\Phi_{n'_1,n'_2}(u,v)du
dv=\delta_{n_1,n'_1}\delta_{n_2,n'_2},
\end{equation}
which is the result of the orthogonality and completeness
properties of the Hermite polynomials. Now, the general solution
of equation (\ref{t}) can be written in the form
\begin{equation}\label{z}
\Psi(u,v)=\sum'_{n_1,n_2}A_{n_1,n_2}\Phi_{n_1,n_2}(u,v),
\end{equation}
where the prime on the sum indicates summation over all values of
$n_1$ and $n_2$ satisfying constraint (\ref{x}). The coefficients
$A_{n_1,n_2}$ are given by \cite{E}
\begin{equation}\label{AB}
\frac{A_{n_1,n_2}}{\sqrt{2^{n_2}n_2!}}=\left(\frac{\pi}{\omega}\right)^{1/4}
\frac{(n_2/2)!c_{n_1}}{(-1)^{n_2/2}n_2!},
\end{equation}
for even values of $n_2$ and are arbitrary for odd values of
$n_2$. In equation (\ref{AB}), $c_n$ is defined as
\begin{equation}\label{AC}
c_n=e^{-1/4|\chi_0|^2}\frac{\chi_0^n}{\sqrt{2^nn!}},
\end{equation}
where $\chi_0$ is an arbitrary complex number. For a more detailed
discussion on the forms of the solutions and their various
graphical representations and different boundary conditions the
reader is referred to \cite{E}.

To understand the predictions of the quantum cosmology of the
model, we note that equation (\ref{t}) is a Schr\"{o}dinger-like
equation for a two dimensional oscillator-ghost-oscillator  with
zero energy moving in the potential
 \begin{equation}\label{AD}
 {\cal U}(u,v)=v^2-u^2-\frac{1}{6}\frac{m}{\omega^2}.
 \end{equation}
 This potential has a saddle point at $u=v=0$, making it possible
 to divide the mini-superspace into two regions characterized by
 ${\cal U}>0$ for the classically forbidden
 or Euclidean and ${\cal U}<0$ for the classically allowed or
 Lorentzian region. The boundary of the two regions is given by
 ${\cal U}=0$, that is
 \begin{equation}\label{AE}
 v^2-u^2=\frac{m}{6\omega^2}=\frac{m}{-\Lambda}.
 \end{equation}
 Comparing the above values of $u$ and $v$ with the classical
 solutions (\ref{e}) and (\ref{f}) shows that they correspond
 to $\beta=0$, thus the same boundary separates the Euclidean
 and Lorentzian regions in both classical and quantum cosmology.
 The behavior of the wave function is exponential in the Euclidean domain
 and oscillatory in the Lorentzian region. Therefore, the smallest
 allowed values for $n_2$ and $n_1$ in (\ref{z}) lead to a wave
 function defined in the Euclidean region. For other values and with more terms,
 equation (\ref{z}) gives the wave function in the Lorentzian domain. As
 in the theory of oscillatory motion in quantum
 mechanics where the square of the wave function has its largest value at the turning points,
 the square of our wave function in the allowed (Lorentzian)
 region has also its largest value corresponding to the values of $u$ and $v$
 satisfying equation (\ref{AE}), {\it i.e.} at the signature
 changing hypersurface in whose vicinity the classical solutions
 are defined. In general, such wavepackets represent a universe with no
singularity at the signature changing hypersurface, that is, the
universe appears from a Euclidean region without tunnelling.
\section{Conclusions}
In this paper we have studied the solutions of Einstein field
equations coupled to a Dirac spinor field in a $4+1$-dimensional
Kaluza-Klein cosmology with a Robertson-Walker type metric and a
negative cosmological constant. We have also explored the
possibility of having solutions that are described by degenerate
metrics signifying transition from a Euclidean to a Lorentzian
domain. The spinor fields representing the matter source in this
study have been taken to be either free or self interacting. In
the case of a free spinor field we obtained  exact solutions for
the resulting Einstein-Dirac system. However, when the matter
field is self interacting, the resulting equations are not
amenable to exact solutions and posses  the property of having
moving singularities. A numerical method for this class of
differential equations has been developed previously and were used
to find the solutions for this case. The behavior of the solutions
shows that the role of the coupling constant in the
self-interacting potential is the same as a perturbation on the
free spinor solutions. Both of these solutions predict signature
transition from a Euclidean to a Lorentzian region. We have shown
that for a very small cosmological constant, signature change can
induce compactification either on the external or internal scale
factors in the Lorentzian region, but with different
compactification scales, that is, $\sim |\Lambda|^{-1/2}$ for $R$
and $\sim |\Lambda|^{1/2}$ for $a$. An interesting property of
this model is that the extra dimension turns out to be stable in
the case of self-interacting spinor fields. The quantum cosmology
of the model for the case when the spinor field is free was also
studied. The corresponding WD equation was found from the
Hamiltonian describing an oscillator-ghost-oscillator system. The
exact solutions of the WD equation show  wavepackets undergoing
signature transition and lead to a quantization condition for the
total energy of the system.

\end{document}